\documentclass[twocolumn,aps,amssymb,showkeys,showpacs,superscriptaddress]{revtex4}
\usepackage{graphicx}
\usepackage{dcolumn}
\usepackage{bm}
\usepackage{psfrag}

\newcommand{\arccot}{\mathop{\rm arccot}\nolimits}

\begin{document}

\title{Ultra-hard fluid and scalar field in the Kerr-Newman metric}

\author{E. Babichev}
\affiliation{APC, Universite Paris 7, rue Alice Domon Duquet,
75205 Paris Cedex 13, France} \affiliation{Institute for Nuclear
Research of the Russian Academy of Sciences, \\ Prospekt 60-letiya
Oktyabrya 7a, Moscow 117312, Russia}
\author{S. Chernov}
\affiliation{Institute for Nuclear Research of the Russian Academy
of Sciences, \\ Prospekt 60-letiya Oktyabrya 7a, Moscow 117312,
Russia} \affiliation{P. N. Lebedev Physical Institute of the
Russian Academy of Sciences,\\ Leninsky Prospekt 53, 119991
Moscow, Russia}
\author{V. Dokuchaev}
\affiliation{Institute for Nuclear Research of the Russian Academy
of Sciences, \\ Prospekt 60-letiya Oktyabrya 7a, Moscow 117312,
Russia}
\author{Yu. Eroshenko}
\affiliation{Institute for Nuclear Research of the Russian Academy
of Sciences, \\ Prospekt 60-letiya Oktyabrya 7a, Moscow 117312,
Russia}

\date{\today}

\begin{abstract}
An analytic solution for the accretion of ultra-hard perfect fluid
onto a moving Kerr-Newman black hole is found. This solution is
a generalization of the previously known solution by Petrich,
Shapiro and Teukolsky for a Kerr black hole.
We show that the found solution is applicable for the case of
a non-extreme black hole,
however it cannot describe the accretion onto an extreme black
hole due to
violation of the test fluid approximation.
We also present a stationary solution for a massless scalar field
in the metric of a
Kerr-Newman naked singularity.
\end{abstract}



\maketitle

\section{Introduction}
%
The only known three-dimensional exact solution for the accretion
flow onto the Kerr black hole is the analytical solution by
Petrich, Shapiro and Teukolsky \cite{PetShapTeu}. This solution
describes the stationary accretion of a perfect fluid with the
ultra-hard equation of state, $p=\rho$, with $p$ being the
pressure and $\rho$ being the energy density, onto a moving Kerr
black hole (see also
\cite{shapiro89,psst89,AbrahamsShapiro90,KarasMucha}). Here we
generalize this solution to the case of a moving Kerr-Newman black
hole. For the Kerr-Newman metric with naked singularity, we
present the stationary solution for a massless scalar field.

The problem of a steady-state accretion onto a (moving) black hole
can be formulated as follows. Consider a black hole moving through
a fluid with a given equation of state, $p=p(\rho)$
 \footnote{Alternatively, one can choose a reference
frame in which the black hole is at rest and the fluid has a
non-zero velocity at the infinity.}. Usually it is assumed that
the back-reaction of fluid to the metric is negligible, in which
case the problem is being solved in the {\sl test fluid
approximation}. It is also assumed that the black hole mass
changes in time slowly enough, such that the steady-state
accretion is established (so called {\sl quasi-stationary}
process). The goal is to find a stationary solution to the
equations of motion for the flowing fluid in the gravitational
field of the black hole.

\begin{figure}[t]
\begin{center}
\includegraphics[width=0.45\textwidth]{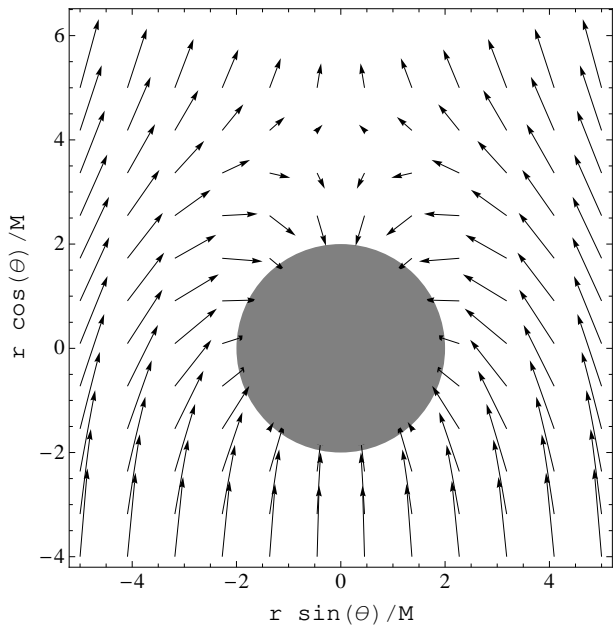}
\end{center}
\caption{\label{arrows}
The three-velocity field of the fluid, ${\bf v}$ for the case of
Schwarzschild black hole, and we take ${\bf v}_{\infty}=0.5$}
\end{figure}

The Fig.~\ref{arrows} illustrates an example of this solution for
the case of fluid passing the Schwarzschild black hole. It is
unlikely that the Kerr-Newman black hole (or a naked singularity)
is found in astrophysical context, as well as the accreting fluid
with the ultra-hard equation of state. However, the theoretical
study of these questions can be useful for better understanding
both the principal General Relativity problems, e.g. the
approaching to the extreme black hole state, and the real matter
accretion onto the astrophysical compact objects.

The velocity of relativistic perfect fluid, $u^\mu$, in the
absence of vorticity can be expressed as the gradient of the
scalar potential $\psi$, see, e.~g.~\cite{LLfluid},
\begin{equation}
 hu_\mu=\psi_{,\mu},
 \label{potential}
\end{equation}
where $h=d\rho/dn=(p+\rho)/n$ is the fluid enthalpy properly
normalized, $h=(\psi^{,\alpha}\psi_{,\alpha})^{1/2}$, $n$ is the
particle number density. When the equation of state for the
perfect fluid has a simple form, $p=p(\rho)$, then the number
density can be expressed in terms of the energy density and the
pressure:
\begin{equation}
 \label{n1}
 \frac{n(\rho)}{n_\infty}=
 \exp\left[\,\,\int\limits_{\rho_{\infty}}^{\rho}
 \frac{d\rho'}{\rho'+p(\rho')}\right],
 \label{numberdens}
\end{equation}
where $n_\infty$ and $\rho_\infty$ are correspondingly the number
density and the energy density at the infinity. The above equation
can be seen as a formal definition of the ``number density'' for a
perfect fluid with an arbitrary equation of state $p=p(\rho)$.

A specific feature of a fluid with the ultra-hard equation of
state is that the resulting equation for the scalar potential is
linear \cite{moncrief}. For $p=\rho$, from Eq.~(\ref{numberdens})
one easily obtains, $\rho\propto n^2$. As a result, the number
density conservation, $(nu^\alpha)_{;\alpha}=0$, gives the
Klein-Gordon equation for a massless scalar field, $\Box\psi=0$.
This reveals a way to obtain an exact solution of the accretion
problem in the Kerr metric by separation of variables and
decomposition in spherical harmonic series \cite{PetShapTeu}. Our
study of the perfect fluids and scalar fields in the Kerr-Newman
metric closely follows the analysis of Petrich {\sl et al.}
\cite{PetShapTeu}.

The introduction of the potential $\psi$, Eq.~(\ref{potential}),
provides a way to solve a problem of hydrodynamics dealing with an
equation for a scalar function. This is not surprising, since
any scalar field $\psi$ is fully equivalent to
the hydrodynamical description, provided that the vector
$\psi^{,\mu}$ is timelike. For example, the canonical massless
scalar field $\psi$ with the energy-momentum tensor,
\begin{equation}
 T_{\mu\nu}=\psi_{,\mu}\psi_{,\nu}-
 \frac{1}{2}g_{\mu\nu}g^{\rho\sigma}\psi_{,\rho}\psi_{,\sigma}.
 \label{energymom}
\end{equation}
is equivalent to a perfect fluid with the ultra-hard equation of
state, provided the following identifications are taken into
account:
\begin{equation}
\label{ident}
 u_\mu \equiv  \frac{\psi_{,\mu}}{\sqrt{\psi_{,\mu}\psi^{,\mu}}},
 \quad p=\rho\equiv \frac{1}{2}\psi_{,\mu}\psi^{,\mu}.
\end{equation}
Therefore, in what follows we do not make a distinction between
massless scalar field and ultra-hard perfect fluid,
as far as vector $\psi^{,\mu}$ is timelike.
For example, the problem of accretion of a perfect fluid can be
viewed as accretion of a scalar field, with a specific form of the
boundary condition at the infinity, $\psi\to \dot{\psi}_\infty t$.
Here $\dot{\psi}_\infty$ is expressed in terms of the fluid
density at the infinity, $\rho_\infty$, as follows,
$\dot{\psi}_\infty=\left(2\rho_\infty\right)^{1/2}$. However, if a
solution for the scalar field contains spacelike $\psi^{,\mu}$ somewhere
then such a solution can be identified with no
perfect fluid analogue.

The paper is organized as follows. Accretion of ultra-hard perfect
fluid (massless scalar field) onto the moving Kerr-Newman black
hole is considered in Sec.~\ref{sec fluid}. The basic equation of
motion for the scalar potential is derived in Sec.~\ref{sec
general}; in Sec.~\ref{sec moving} we solve this equation for the
case of the moving non-extreme black hole; in Sec.~\ref{sec
extreme} we analyze in detail the case of the extreme black hole.
In Sec.~\ref{sec sf} we find a stationary solution for the scalar
field in the Kerr-Newman singularity. In Sec.~\ref{sec conclusion}
we briefly describe the results of the work.

\section{Accretion of ultra-hard fluid}
\label{sec fluid}
\subsection{Metric and equation of motion}
\label{sec general}

The Kerr-Newman metric can be written as \cite{mtw,Gal'tsov},
\begin{equation}
 ds^2=\frac{\Sigma\Delta}{\mathcal A}dt^2
 -\frac{{\mathcal A}\sin^2\theta}{\Sigma}(d\phi-\omega dt)^2-
 \frac{\Sigma}{\Delta}dr^2-\Sigma d\theta^2,
 \label{metric}
\end{equation}
where
\begin{eqnarray}
\Delta &= & r^2-2Mr+a^2+Q^2; \label{Delta} \\
\Sigma &=& r^2+a^2\cos^2\theta;  \label{Sigma} \\
 {\mathcal A} &=& (r^2+a^2)^2-a^2\Delta\sin^2\theta; \label{A} \\
 \omega &=& \frac{2Mr-Q^2}{\mathcal A}\, a.
 \label{omega}
\end{eqnarray}
Here $M$ is the mass of a black hole or a naked singularity, $a$
is the specific angular momentum, $Q$ is the electric charge and
$\omega$ is the angular dragging velocity. The event horizon of
the Kerr-Newman black hole, $r=r_+$, is the larger root of the
equation $\Delta=0$, i.~e. $r_{\pm}=M\pm\sqrt{M^2-a^2-Q^2}$. The
event horizon exists only if $M^2\geq a^2+Q^2$. When $M^2<a^2+Q^2$
the metric (\ref{metric}) describes a naked singularity.

The Klein-Gordon equation for a massless scalar field in the
background gravitational field $g_{\alpha\beta}$ is given by
\begin{equation}
\psi^{;\alpha}_{\,\,\,\,;\alpha}=\frac{1}{\sqrt{-g}}\frac{\partial
}{\partial
x^\alpha}\left(\sqrt{-g}g^{\alpha\beta}\frac{\partial\psi}{\partial
x^\beta}\right)=0.
\end{equation}
For the Kerr-Newman metric (\ref{metric}) the above equation can
be rewritten as
\begin{widetext}
\begin{equation}
\left\{\frac{1}{\Delta}\!\left[(r^2\!+\!a^2)\partial_t\!
 +\!a\partial_\phi\right]^2\!-\frac{1}{\sin^2\theta}\left(\partial_\phi\!
 +\!a\sin^2\theta\partial_t\right)^2\!-
 \partial_r\!\left(\Delta\partial_r\right)
 -\frac{1}{\sin\theta}\partial_\theta
 (\sin\theta\,\partial_\theta)\right\}\psi\!=\!0.
 \label{equat}
\end{equation}
\end{widetext}
Eq.~(\ref{equat}) is the main equation we will use to find
analytic solutions.

\subsection{Accretion onto moving black hole}
\label{sec moving}
%
First we consider a stationary accretion of an ultra-hard perfect
fluid onto a non-extreme black hole, i.~e. we assume $M^2>
a^2+Q^2$. The first boundary condition for Eq.~(\ref{equat}) is a
relation at the space infinity \cite{PetShapTeu},
$r\rightarrow\infty$:
\begin{eqnarray}
  \label{bound1}
 \psi&=&-u_\infty^0 t +u_\infty r
 [\cos\theta\cos\theta_0\nonumber \\
 &+& \sin\theta\sin\theta_0\cos(\phi-\phi_0)],
\end{eqnarray}
where $u_\infty^0$ is the 0-component of the black hole 4-velocity
at infinity
\begin{equation}
 u^\mu\equiv (u^0,{\mathbf u}_\infty)=(1-v^2)^{-1/2}
 (1,{\mathbf v}_\infty),
\end{equation}
and $u_\infty\equiv |{\mathbf u}_\infty|$; while ${\mathbf
v}_\infty$ is the 3-velocity vector of the black hole at the
infinity with the space orientation specified by the two arbitrary
angles, $\theta_0$ and $\phi_0$.

\begin{figure}[t]
\begin{center}
\includegraphics[width=0.45\textwidth]{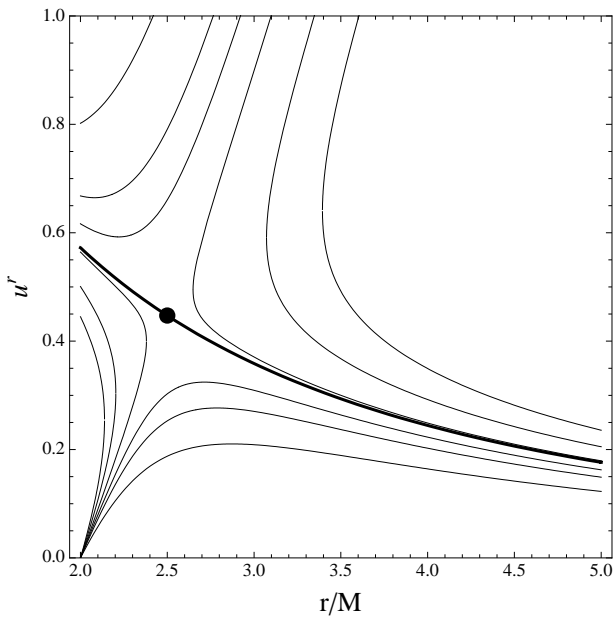}
\end{center}
\caption{\label{figcrit} Possible types of solutions for the
radial component of fluid four-velocity are shown as a function of
the radius for the accretion of fluid with equation of state
$p=\rho/2$ onto a Schwarzschild black hole. Different curves
correspond to different values of the influx. The single solution
with a critical value of inflow, corresponding to the physical
solution, is shown by the solid curve with a dot labeling the critical point.}
\end{figure}

There are several approaches to formulate the second boundary
condition for Eq.~(\ref{equat}) in the case of the stationary
accretion onto black hole.

The classical approach, originated from the Bondi problem
\cite{bondi,zeldovich}, fixes the value of the energy flux onto a
black hole at the critical sound surface
\cite{Michel,Beskin97,Beskin04,Bogovalov}. The critical point is
fixed by the requirement, that the physical solution is double
valued neither in the velocity of the fluid nor in the radial
coordinate. This requirement picks up one physical solution from
the infinite number of formal solutions, parameterized by the flux
of the energy onto a black hole. In Fig.~\ref{figcrit} several
solutions with different fluxes are shown, along with the
``correct'' (critical) one.

An alternative way to fix the flux is to ask that the energy
density is finite at the black hole event horizon $r=r_+$ (see
e.g. \cite{PetShapTeu}). In most cases both approaches give the
same result. In particular, this is true for a non-extreme
Kerr-Newman black hole. As we see later, however, in the specific
case of the extreme Kerr-Newman black hole the requirement for the
flow to have a trans-sonic point gives a solution for which the
energy density is infinite at the horizon (see details in the next
Section~\ref{sec extreme}).

Note, that at the critical surface the infall velocity of the
accreting fluid reaches the sound speed $c_s$, which for the
ultra-hard fluid is equal to the speed of light, $c_s=1$. For this
reason in the considered case a sound surface coincides with the
event horizon $r=r_+$.

Following Petrich {\sl et al.} \cite{PetShapTeu}, we search a
solution of Eq.~(\ref{equat}) separating variables and decomposing
$\psi$ in the spherical harmonic series:
\begin{equation}
 \psi=-u^0_\infty t+\sum_{l,m}A_{l,m}R_l(r)Y_{lm}(\theta,\phi),
 \label{solut}
\end{equation}
where the radial part $R_l(r)$ satisfies the equation
\begin{equation}
 \frac{d}{d r}\left[\Delta\frac{d R_l(r)}{d r}\right]+
 \left[-l(l+1)+\frac{m^2a^2}{\Delta}\right]R_l(r)=0.
 \label{radial}
\end{equation}
It is convenient to define a new variable $\xi$ by the relation,
\begin{equation}
\label{redef} r=M+\xi\sqrt{M^2-a^2-Q^2}.
\end{equation}
Note that the definition of $\xi$, Eq.~(\ref{redef}), does not
work in the case of the extreme black hole. Using (\ref{redef}) we
can rewrite Eq.~(\ref{radial}) in the form
\begin{equation}
 (1-\xi^2)R^{''}_{\xi\xi}-2\xi R^{'}_\xi+\left[l(l+1)-
 \frac{m^2(i\alpha)^2}{1-\xi^2}\right]R=0,
 \label{Legandre}
\end{equation}
where $\alpha=a/\sqrt{M^2-a^2-Q^2}$. This is the Legendre equation
with an {\it imaginary} second index \cite{abram}. The general
solution of Eq.~(\ref{solut}) is
\begin{eqnarray}
 \psi&\!\!\!=&\!\!\!-u^0_\infty t+\sum_l[A_lP_l(\xi)\!+
 \!B_lQ_l(\xi)]Y_{l0}(\theta,\phi) \nonumber \\
 &\!\!\!+&\!\!\!\sum_{lm}{}^{'}[A_{lm}^{+}P_l^{im\alpha}(\xi)\!+\!
 A_{lm}^{-}P_l^{-im\alpha}(\xi)]Y_{lm}(\theta,\phi).
  \label{GenSol}
\end{eqnarray}
In the above equation the prime $(')$ denotes that we do not
include the term with $m=0$; and $P_l$ and $Q_l$ are the Legendre
functions of the first and second kind correspondingly; and
$P_l^{im\alpha}$ is the associated Legendre function which can be
expressed in terms of the hypergeometric function \cite{abram}:
\begin{equation}
 P_l^{im\alpha}(\xi)\propto e^{im\chi}F(-l,l+1;1-im\alpha;(1-\xi)/2)
 \label{hypergeom}
\end{equation}
where
\begin{equation}
 \chi=\frac{\alpha}{2}\ln\frac{\xi+1}{\xi-1}=
 \frac{a}{2\sqrt{M^2-a^2-Q^2}}\ln\frac{r-r_{-}}{r-r_+}.
 \label{chi}
\end{equation}
Then the components of the 4-velocity are
\begin{widetext}
\begin{eqnarray}
 nu_t &=& -u^0_\infty \nonumber\\
 nu_r&=&\frac{\big\{\sum_{l}[A_l(P_l)_\xi^{'}+B_l(Q_l)^{'}_\xi]Y_{l0}
 +\sum^{'}_{lm}[A^+_{lm}(P_{l}^{im\alpha})^{'}_{\xi}
 +A_{lm}^-(P_{l}^{-im\alpha})^{'}_\xi]Y_{lm}\big\}}{\sqrt{M^2-a^2-Q^2}};
 \nonumber\\
 nu_\theta&=&\left\{\sum_l[A_lP_l+
 B_lQ_l]\frac{\partial Y_{l0}}{\partial\theta}+\sum_{lm}{}^{'}[A^
 +_{lm}P_{l}^{im\alpha}+A_{lm}^-P_{l}^{-im\alpha}]\frac{\partial
 Y_{lm}}{\partial\theta}\right\}; \nonumber\\
 nu_\phi&=&\left\{\sum_{lm}{}^{'}\left[A^+_{lm}P_{l}^{im\alpha}
 +A_{lm}^-P_{l}^{-im\alpha}\right]\frac{\partial
 Y_{lm}}{\partial\phi}\right\},
\end{eqnarray}
where subindex $\xi$ denotes a derivative on the variable $\xi$.
Using the normalization condition $u_\alpha u^\alpha=1$, we obtain
\begin{equation}
 n^2=(\Sigma\Delta)^{-1}[(r^2+a^2)u^0_\infty-anu_\phi]^2-
 (\Sigma\sin^2\theta)^{-1}[nu_\phi-a\sin^2\theta u^0_\infty]^2
 -\frac{\Delta}{\Sigma}(nu_r)^2-\Sigma^{-1}(nu_\theta)^2.
 \label{n}
\end{equation}
In the limit $r\to r_+$ we find,
\begin{eqnarray}
 (n^2\Delta\Sigma)|_{r\to r_+}&=&\big[(r^2_++a^2)u_\infty^0
 -\sum_{l,m}{}^{'}im(A^+_{lm}e^{im\chi}+
 A^-_{-lm}e^{-im\chi})Y_{lm}(\theta,\phi)\big]^2 \\
  &-&\big[a\sum_{l,m}{}^{'}im(-A^+_{lm}e^{im\chi}
 +A^-_{lm}e^{-in\chi})Y_{lm}(\theta,\phi)-\sqrt{M^2-a^2-Q^2}
 \sum_{l}B_lY_{lm}(\theta,\phi)\big]^2\!\!.\nonumber
\end{eqnarray}
From the finiteness of the fluid density at the event horizon it
follows that all $B_l$ equal to zero except $B_{0}$, and
$B_{0}>0$. All coefficients $A_{lm}^+$ are also zero, since the
terms containing $A_{lm}^+$ are irregular at the horizon. The
corresponding solution, which is finite at the horizon, reduces to
\begin{equation}
 \psi = -u_\infty^0t+\frac{(r^2_+
 +a^2)u_\infty^0}{2\sqrt{M^2-a^2-Q^2}}\ln{\frac{r-r_-}{r-r_+}}
 +\sum_{l,m}A_{lm}F(-l,l+1;1+im\alpha;(1-\xi)/2)Y_{lm}(\theta,\phi-\chi),
\end{equation}
where the second term comes from the Legendre function $Q_0(x)$:
\begin{equation}
 Q_0(x)=\frac{1}{2}\ln{\frac{1+x}{1-x}}.
\end{equation}
Using the boundary condition at the infinity, Eq.~(\ref{bound1}),
we finally obtain,
\begin{eqnarray}
 \label{PsiFinBlackHole}
 \psi=-u_\infty^0t+\frac{(r^2_++a^2)u_\infty^0}
 {2\sqrt{M^2-a^2-Q^2}}\ln{\frac{r-r_-}{r-r_+}}
 +u_\infty(r\!-\!M)\cos\theta\cos\theta_0
 +u_\infty \mbox{Re}[(r\!-\!M\!+\!ia)\sin\theta\sin\theta_0
 e^{i(\phi-\phi_0-\chi)}].
\end{eqnarray}
In the case of the Kerr black hole the solution
(\ref{PsiFinBlackHole}) reduces to the known solution by Petrich
{\sl et al} \cite{PetShapTeu}. From (\ref{PsiFinBlackHole}) one
can find the components of the 4-velocity:
\begin{eqnarray}
 \qquad nu_t &=& -u_\infty^0,\nonumber\\
 nu_r &=&-u_\infty^0\frac{r^2_+\!+\!a^2}{\Delta}
 \!+\!u_\infty\cos\theta\cos\theta_0
 \!+\!u_\infty \mbox{Re}\{[1\!+\!\frac{ia}{\Delta}(r\!-\!M\!+
 \!ia)]\sin\theta\sin\theta_0e^{i(\phi-\phi_0-\chi)}\}; \nonumber\\
  nu_\theta&=&-u_\infty(r-M)\sin\theta\cos\theta_0+u_\infty
 \mbox{Re}[(r-M+ia)\cos\theta\sin\theta_0 e^{i(\phi-\phi_0-\chi)}];
 \nonumber\\
  nu_\phi &=& -u_\infty \mbox{Im}[(r-M+ia)\sin\theta\sin\theta_0
 e^{i(\phi-\phi_0-\chi)}].
\end{eqnarray}
\end{widetext}
The accretion rate of the fluid onto the black hole is given by
\begin{equation}
 \dot{N}\!=\!-\!\int_S nu^i\sqrt{-g}\,dS_I
    =4\pi(r^2_++a^2)n_\infty u^0_\infty.
 \label{flux}
\end{equation}
While, the corresponding rate of the energy flux onto the black
hole is
\begin{equation}
 \dot{M}=4\pi(r^2_++a^2)(\rho_\infty+p_\infty)u^0_\infty.
 \label{dotM}
\end{equation}
Note that as in the Kerr case, the flux (\ref{flux}) is
independent on the direction of the black hole motion.

From Eq.~(\ref{PsiFinBlackHole}) one can see, that in the limit of
the extreme black hole, $M^2-a^2-Q^2\to 0$, the solution for the
accretion is problematic, since $\psi$ contains a divergent term
proportional to $\left(M^2-a^2-Q^2\right)^{-1/2}$. Moreover, in
this limit $n\to \infty$ while $u^r\to 0$ at the event horizon,
which clearly signals on the violation of the test fluid
approximation. However, it might happen, that going to the limit
$M^2-a^2-Q^2\to 0$ we missed a ``correct'' regular solution, valid
only in the extreme case. To check this, in the next Section
\ref{sec extreme}, we will redo all the calculation for the
extreme case, $M^2=a^2+Q^2$.

\begin{figure}[t]
\begin{center}
\includegraphics[width=0.45\textwidth]{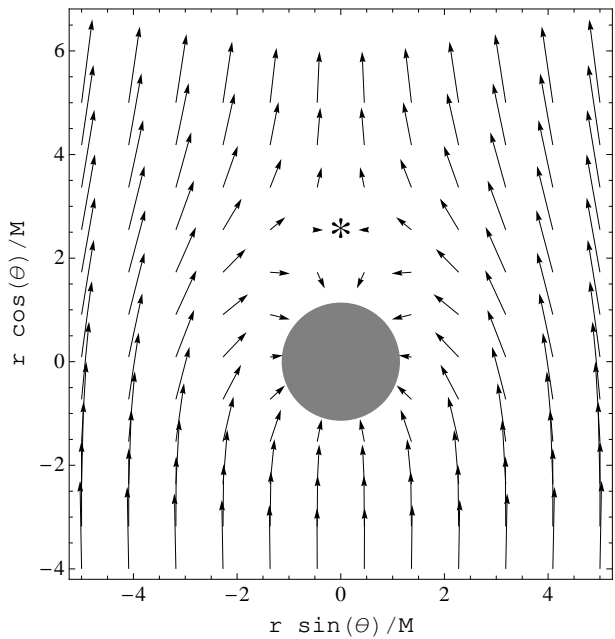}
\end{center}
\caption{\label{arrows1}
The three-velocity of the fluid in the case of almost extremal
Reissner-Nordstr\"om black hole, $Q=0.99 M$. We take ${\bf
v}_{\infty}=0.5$. The stagnation point is indicated by star.}
\end{figure}

Let us study in more details some particular cases. First, we set
$a=0$, i.~e. we consider a moving non-rotating charged black hole.
As in the case of a Schwarzschild black hole (see
\cite{PetShapTeu}), there is a stagnation point: the point where
the fluid velocity is zero relative to the black hole. It is not
difficult to find the location of the stagnation point:
\begin{eqnarray}
\label{stagnation}
r_{stag} = M\left\{1+\left(1+\frac{2r_+M-Q^2(1+v_\infty)}
{M^2v_\infty}\right)^{1/2}\right\}\nonumber
\end{eqnarray}
As an example, in the Fig.~\ref{arrows1} we show the
three-velocity of the fluid in the case of Reissner-Nordstr\"om
black hole with $Q=0.99 M$ and ${\bf v}_{\infty}=0.5$.

\begin{figure}[t]
\begin{center}
\includegraphics[width=0.45\textwidth]{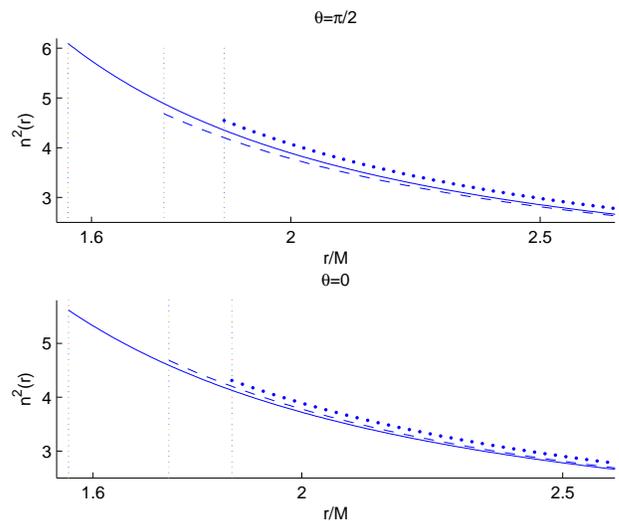}
\end{center}
\caption{\label{BlacHol} Radial density distribution of the
stationary accreting ultra-hard fluid around the Kerr-Newman black
hole is shown in the equatorial plane ($\theta=\pi/2$) and along
the polar axis ($\theta=0$). In both plots the solid lines
correspond to the case $Q=(2/3)M$, $a=M/2$, dashed lines are for
$Q=(2/3)M$, $a=0$ and dotted line are for $Q=0$, $a=M/2$
respectively.}
\end{figure}

Let us turn to another particular case. Substituting $u_\infty=0$
into Eq.~(\ref{n}) we find the radial density distribution of the
accreting ultra-hard fluid onto the Kerr-Newman black hole at
rest:
\begin{equation}
 n^2=\frac{{\mathcal A}-(r_+^2+a^2)^2}{\Sigma\Delta}.
 \label{densityaccr}
\end{equation}
This radial distribution is shown in Fig.~\ref{BlacHol}. At the
event horizon $r=r_+$ the ratio of densities at the equator
($\theta=\pi/2$) and the pole ($\theta=0$) is
\begin{equation}
 \frac{\rho(r_+,\pi/2)}{\rho(r_+,0)}=
 \frac{4r_+(r^2_+ +a^2)-a^2(r_+-r_-)}{4r_+^3}.
\end{equation}
In Appendix~\ref{sec rest} we provide an alternative approach to
solve a problem for the stationary accretion, in the case when a
black hole is at rest.
%
\subsection{Accretion onto extreme black hole}
\label{sec extreme}
%
In this section we will concentrate specifically on the accretion
onto the extreme Kerr-Newman black hole, trying to find a good
solution. We search a stationary solution of the wave equation
(\ref{equat}) in the form (\ref{solut}), like in the non-extreme
case. The radial part $R_l(r)$ now satisfies Eq.~(\ref{radial})
with $\Delta=(r-M)^2$. Using a new variable,
$$\zeta=r/M-1,$$
we transform (\ref{radial}) to a simpler equation
\begin{equation}
 \zeta^2 R^{''}_{\zeta\zeta}+2\zeta R_{\zeta}^{'}
 +\left[\frac{m^2a^2}{M^2\zeta^2}-l(l+1)\right]R=0.
 \label{ExtrWavEquaR}
\end{equation}
For $m=0$ Eq.~(\ref{ExtrWavEquaR}) reduces to the Euler's equation
with a solution
\begin{equation}
 R=C_1\left(\frac{r}{M}-1\right)^l+C_2\left(\frac{r}{M}-1\right)^{-l-1}.
\end{equation}
The corresponding potential is
\begin{eqnarray}
 \psi &=& -u^0_\infty t\\
    &+&\sum_{l,m}\left[C^1_{lm}\left(\frac{r}{M}-1\right)^l
    +C^2_{lm}\left(\frac{r}{M}-1\right)^{-l-1}\right]Y_{lm}.\nonumber
\end{eqnarray}
Using the boundary condition at infinity we find:
\begin{equation}
 \psi=-u_\infty^0t+\frac{u_\infty^0M^2}{r-M}+u_\infty (r-M)\cos\theta.
\end{equation}
For $m\neq0$ the solution of Eq.~(\ref{radial}) is
\begin{equation}
 R\!=\!\sqrt{\frac{ma}{M\xi}}\left[C_1J_{l+1/2}
 \left(\frac{ma}{M\xi}\right)\!+\!
 C_2\Upsilon_{l+1/2}\left(\frac{ma}{M\xi}\right)\right]\!\!,
\end{equation}
where $J$ and $\Upsilon$ is Bessel function of the first and
second kind correspondingly. The general solution of
Eq.~(\ref{ExtrWavEquaR}) is
\begin{widetext}
\begin{eqnarray}
 \psi &=& -u_\infty^0 t
    +\sum_{l}\left[C^1_{l0}\left(\frac{r-M}{M}\right)^l
    +C^2_{l0}\left(\frac{r-M}{M}\right)^{-l-1}\right]Y_{l0}
    \nonumber     \\
    &&+\sum_{l,m}{}^{'}\sqrt{\frac{ma}{M\xi}}\left[C^{1+}_{lm}J_{l+1/2}
    \left(\frac{ma}{M\xi}\right)
    +C^{2-}_{lm}\Upsilon_{l+1/2}\left(\frac{ma}{M\xi}\right)\right]Y_{lm},
\end{eqnarray}
where it used the Bessel functions \cite{PolZai}:
\begin{eqnarray}
 J_\nu(x)=\sum_{k=0}^\infty\frac{(-1)^k(x/2)^{\nu+2k}}{k!\Gamma(\nu+k+1)},
  \\
 \Upsilon_{\nu}(x)=\frac{J_\nu(x)\cos\pi\nu -J_{-\nu}(x)}{\sin\pi\nu}, \\
 J_{-3/2}(x)=\sqrt{\frac{2}{\pi x}}\left(-\frac{\cos x}{x}-\sin x\right).
\end{eqnarray}
From the first boundary condition at space infinity we find
$C^1_{l0}=C_{1m}^{2-}=0$, $C_{10}^1=Mu_\infty\cos\theta_0$,
$C_{11}^{2-}=-\sqrt{\pi/2}u_\infty a\sin\theta_0$. While, the
second boundary condition fixes the value of the influx at the
sound surface: $C^{1+}_{lm}=C^2_{l0}=0$,
$C_{00}^2=u^0_\infty(M^2+a^2)/M$. With these boundary conditions
we obtain
\begin{eqnarray}
 \psi &=& -u_\infty^0 t+u_\infty(r\!-\!M)\cos\theta\cos\theta_0
 +\frac{(M^2+a^2)u^0_\infty}{r\!-\!M} \nonumber\\
 && +u_\infty a\sin\theta\sin\theta_0\cos(\phi-\phi_0)\!\left(\frac{r\!
 -\!M}{a}\cos\frac{a}{r\!-\!M}+\sin\frac{a}{r\!-\!M}\right).
 \label{potentialextreme}
\end{eqnarray}
The components of the 4-velocity are given by
\begin{eqnarray}
 nu_t &=& -u_\infty^0, \nonumber\\
 nu_r &=&-\frac{(M^2+a^2)u_\infty^0}{(r\!-\!M)^2}
 +u_\infty\cos\theta\cos\theta_0 \nonumber \\
 &+&u_\infty\sin\theta_0\sin\theta\cos(\phi-\phi_0)
 \cos\left(\frac{a}{r-M}\right)\left[1-\frac{a^2}{(r\!-\!M)^2}+
 \frac{a}{r\!-\!M}\tan\left(\frac{a}{r-M}\right)\right], \nonumber\\
  nu_\theta&=&-u_\infty(r-M)\sin\theta\cos\theta_0
 +u_\infty a\cos\theta\sin\theta_0
 \cos(\phi-\phi_0)\left[\frac{r\!-\!M}{a}\cos\frac{a}{\!r\!-M}
 +\sin\frac{a}{r\!-\!M}\right],\nonumber\\
  nu_\phi&=&-u_\infty a \sin\theta_0\sin\theta\sin(\phi-\phi_0)
 \left[\frac{r\!-\!M}{a}\cos\frac{a}{r\!-\!M}+\sin\frac{a}{r\!-\!M}\right].
 \label{4velocity}
\end{eqnarray}
\end{widetext}
Using the above solution, one can calculate the accretion rate
\begin{eqnarray}
 \dot{N}=4\pi(M^2+a^2)u^0_\infty n_\infty,
 \label{ratextreme}
\end{eqnarray}
and radial density distribution (for $u_\infty=0$)
\begin{eqnarray}
 n^2=\frac{(r^2+a^2)^2-(M^2+a^2)^2}{\Sigma(r-M)^2}
 -\frac{a^2\sin^2\theta}{\Sigma}.
 \label{densextreme}
\end{eqnarray}
From (\ref{4velocity}) and (\ref{densextreme}) one can see that at
the event horizon of the extreme black hole $r_+=M$ both the
radial 4-velocity $u^r$ and the energy density $\rho$  behave
as: $u^r\to 0$ and $\rho\propto
n^2\propto(r-M)^{-1}\to\infty$. And the full mass of fluid near
the black hole tends to infinity too. This behavior is an
indication of violation of the test fluid approximation. For this
reason the solution (\ref{potentialextreme}), (\ref{4velocity})
and (\ref{densextreme}) is not self-consistent, and back reaction
of the accreting fluid must be taken into account to obtain a
physically relevant solution.
%
\section{Scalar field around naked singularity}
\label{sec sf}
%
It is worthwhile to note that the spacetime of the Kerr-Newman
naked singularity is pathological when both $Q$ and $a$ are
nonzero \cite{carter68}. In general, this spacetime is not
globally hyperbolic, which means, in particular, that the Cauchy
problem for the scalar field is not well-posed. The origin of the
pathology is inside the region $\mathcal{A}<0$, where function
$\mathcal{A}$ is given by Eq.~(\ref{A}). This region contains
closed causal curves, and it is possible to start from a point at
the asymptotically flat region far from the singularity, to go
into the ``pathological'' region and to travel back in time there
and then come back to the starting point. Thus a time machine can
be constructed. If, however, $Q=0$ or $a=0$ the space-time is not
pathological anywhere, except for the point of the physical
singularity at $r=0$.

Below we describe an analytic solution for the stationary
distribution of the scalar field in the Kerr-Newman geometry. For
simplicity we consider the case when the naked singularity is at
rest, with respect to the scalar field at the infinity. Thus we
impose the boundary condition at the infinity in the following
form,
\begin{equation}
 \label{bound3}
 \psi\to\dot{\psi}_\infty t+\psi_\infty \quad \mbox{at}
 \quad r\to\infty,
\end{equation}
where $\dot{\psi}_\infty$ and $\psi_\infty$ are constants.

\begin{figure}[t]
\begin{center}
\includegraphics[width=0.45\textwidth]{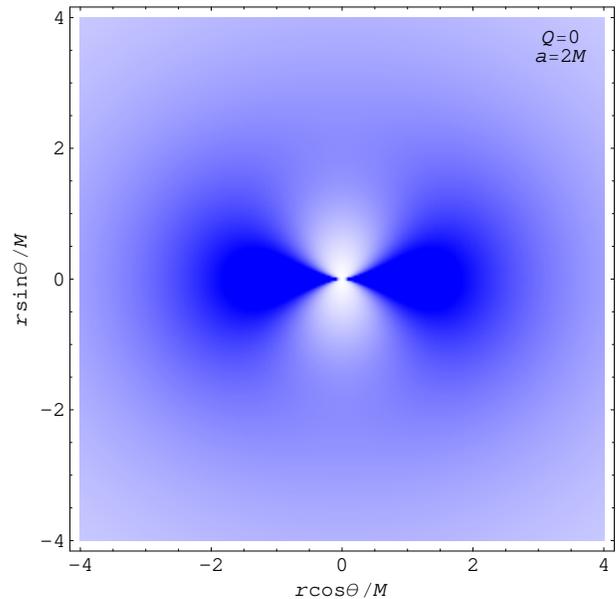}
\end{center}
\caption{\label{NakSing2} Density plot in the poloidal plane of
the energy density distribution of ultra-hard perfect fluid
(massless scalar field) around the Kerr naked singularity with
$a=2M$. The particular solution (\ref{scalar2}) with
$\dot\psi_\infty\neq0$ and $B=0$ is taken, so that the influx onto
the singularity is zero.}
\end{figure}

The corresponding solution of the wave equation (\ref{equat}) is
given by Eq.~(\ref{GenSol}), provided that the following
replacements are made,
$$\xi\to-i\xi,\quad \alpha\to i\alpha,\quad
-u_\infty^0\to\dot\psi_\infty.$$
The boundary condition at the
infinity gives $A_{lm}^+=A_{lm}^-=0$ for all $l$ and $m$, since
the solution should not depend on $\phi$ as $r\to\infty$. To
satisfy Eq.~(\ref{bound3}) it is also necessary to put $A_l=0$ and
$B_l=0$ for all $l\neq 0$, since the terms corresponding $l\neq 0$
contain solutions which diverge at $r\to\infty$. The term
containing $A_0$ is a constant which can be identified with
$\psi_\infty$. The full solution which satisfies the boundary
condition (\ref{bound3}) can be written as
\begin{equation}
 \psi=\dot{\psi}_\infty t+\psi_\infty+\!
 B\!\left[\arctan\!\left(\frac{r-M}{\sqrt{a^2\!
 +\!Q^2\!-\!M^2}}\right)\!-\!\frac{\pi}{2}\!\right]\!,
 \label{scalar2}
\end{equation}
where the last term comes from $Q_0$, and $B$ is an arbitrary
constant. The components of the energy-momentum tensor are:
\begin{eqnarray}
 \sqrt{-g}\,\,T_t^r
 &=&-\dot{\psi}_\infty B\sin\theta\sqrt{a^2+Q^2-M^2},
 \label{scalarflux} \\
 T_t^t=-T_r^r&=&\frac{B^2(a^2+Q^2-M^2)+
 {\mathcal A}\dot{\psi}_\infty^2}{2\Sigma\Delta},
 \label{scalarT} \\
 T_\theta^\theta=T_\phi^\phi &=&\frac{B^2(a^2+Q^2-M^2)-
 {\mathcal A}\dot{\psi}_\infty^2}{2\Sigma\Delta}.
\end{eqnarray}
The energy flux onto the singularity is nonzero when
$B\dot{\psi}_\infty\neq0$. The corresponding radial 4-velocity of
the inflowing fluid is
\begin{equation}
 u^r=-\frac{B\sqrt{a^2+Q^2-M^2}\,\Delta}
 {\dot{\psi}_\infty^2{\mathcal A}-B^2(a^2+Q^2-M^2)}.
\end{equation}
Note that the influx (\ref{scalarflux}) is not fixed when both
constants $B$ and $\dot{\psi}_\infty$ are nonzero, because in the
case of the naked singularity one physical boundary condition is
missing, which for the black hole was the boundary condition at
the event horizon or at the critical point. The energy flux onto
the central singularity is zero in particular cases, $B=0$ or
$\dot{\psi}_\infty=0$.

\begin{figure}[t]
\begin{center}
\includegraphics[width=0.45\textwidth]{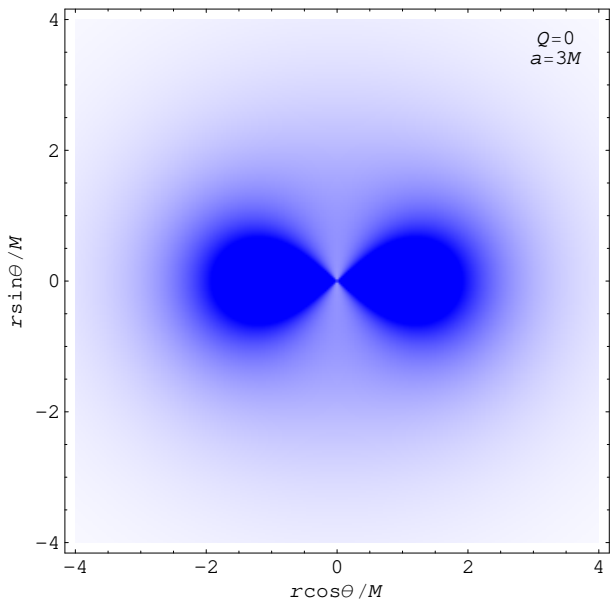}
\end{center}
\caption{\label{NakSing3} Density plot in the poloidal plane of
the energy density distribution of a scalar field around the Kerr
naked singularity with  $a=3M$. A solution (\ref{scalar2}) with
$\dot\psi_\infty=0$ and $B\neq 0$ is taken, and the influx onto
the singularity is zero. This solution does not have a perfect
fluid analogue, and the total mass of the scalar field for this
solution is finite.}
\end{figure}

A particular solution (\ref{scalar2}) with $B=0$ and
$\dot\psi_\infty\neq0$ describes the stationary distribution of
the massless scalar field (or, equivalently, the ultra-hard fluid
with $p=\rho$) without the influx around the naked singularity.
This solution has a very specific feature: the energy density of
the scalar field $\rho(r,\theta)=T_t^t$ from Eq.~(\ref{scalarT})
becomes zero at the surface $\mathcal{A}=0$, inside of which the
causality is violated.

Since the space-time with both $a$ and $Q$ non-zero is
pathological, in what follows we consider only the Kerr and
Reissner-Nordstr\"om cases, correspondingly $Q=0$ and $a=0$, where
noncausal region $\mathcal{A}<0$ is absent. For example, in
Fig.~\ref{NakSing2} the distribution of the energy density of
ultra-hard fluid with zero flux around the Kerr naked singularity
is shown.

An another particular solution (\ref{scalar2}) with
$\dot\psi_\infty=0$ and $B\neq0$, describes a stationary
distribution of the massless scalar field with zero energy density
at infinity. This solution does not have a perfect fluid analogue.
In Fig.~\ref{NakSing3} the energy density distribution for the
scalar field is shown. The total mass of the scalar field for this
solution is finite, $M_f<\infty$. To calculate this mass we use
the quasi-static coordinate frame $(t,r,\theta,\tilde\phi)$ with
azimuthal coordinate $\tilde\phi$,
\begin{equation}
 d\tilde\phi=d\phi-\omega dt.
  \label{tildephi}
\end{equation}
This coordinate frame is ``rotating with the geometry''
\cite{bardeen1,bardeen2} with the angular velocity $\omega$ given
by (\ref{omega}). An observer at rest in this frame rotates in the
azimuthal direction with an angular velocity $\omega$. The total
mass of the scalar field stationary distributed around the naked
singularity is
\begin{equation}
 M_f=\int\xi_{(t)}^\alpha T_\alpha^\beta\sqrt{-g}\,d\Sigma_\beta
 =\int\tilde T_t^t\sqrt{-\tilde g}\,d^{\,3}\tilde x,
 \label{massf}
\end{equation}
where $d^{\,3}\tilde x=drd\theta d\tilde\phi$ and $\sqrt{-\tilde
g}=\Sigma\sin\theta$. Using expression for the energy-momentum
tensor (\ref{energymom}), we calculate the scalar energy density
$\tilde T_t^t$ for the solution (\ref{scalar2}) in the
quasi-static frame $(t,r,\theta,\tilde\phi)$:
\begin{equation}
 \tilde T_t^t=-\frac{1}{2}\,\tilde g^{rr}(\psi_{,r})^2
 =\frac{1}{2}B^2\frac{\epsilon^2}{\Sigma\Delta}M^2,
 \label{tildeT00}
\end{equation}
where we define a super-extreme parameter
\begin{equation}
 \epsilon=\sqrt{\frac{a^2+Q^2-M^2}{M^2}}, \quad a^2+Q^2-m^2>0.
 \label{epsilon}
\end{equation}
Substituting $\tilde T_t^t$ from (\ref{tildeT00}) into
(\ref{massf}), we obtain
\begin{equation}
 M_f=2\pi B^2\epsilon^2M^2\int\limits_0^\infty\frac{dr}{\Delta}
 =\pi B^2(\pi+2\arccot\epsilon)\,\epsilon M.
\end{equation}
This equation relates the constant $B$ in (\ref{scalar2}) with a
total mass of the scalar field.

\section{Conclusion}
\label{sec conclusion}
We found an analytic solution for the stationary accretion of
ultra-hard perfect fluid onto the moving Kerr-Newman black hole.
The presented solution is a generalization of Petrich, Teukolsky
and Shapiro solution for the accretion onto the moving Kerr black
hole. Our solution describes the velocity, and the density
distribution of the perfect fluid with the equation of state
$p(\rho)=\rho$ in space, when the Kerr-Newman black hole moves
through the fluid. The solution is unequally determined in terms
of the scalar potential, the exact solution for which is given by
Eq.~(\ref{PsiFinBlackHole}). One can think of the found solution
as of one to the problem for a canonical massless scalar field in the
metric of a moving Kerr-Newman black hole, since these two
description are equivalent for the considered problem. In this
case the solution for the scalar field coincides with the scalar
potential for the fluid, Eq.~(\ref{PsiFinBlackHole}).

Our results are only valid in the test fluid approximation, put
differently, the back reaction of the accreting fluid is ignored.
In the case of a non-extreme Kerr-Newman black hole one can always
find such a range of parameters that the back reaction can indeed
be safely ignored. While this is not the case for the extreme
black hole. We showed that for the extreme Kerr-Newman black hole
the test fluid approximation is always violated, since the energy
density of ultra-hard fluid diverges at the event horizon. A
similar behavior of the energy density for the perfect fluids with
sound velocity $c_s\geq1$ occurs for the extreme
Reissner-Nordstr\"om black hole \cite{bcde08a}. Apparently, if the
black hole is almost extreme, $M^2-Q^2-a^2\to +0$, the value of
the energy density at the infinity must be tuned considerably, to
avoid the violation of the test fluid approximation. Thus to solve
correctly the problem of accretion in the case of an (almost)
extreme black hole the back reaction of the fluid must be taken
into account. This is in accord with \cite{hod02,hod08}, where it
was suggested that the back reaction is important for a
description of a scalar particle absorption with a large angular
momentum by a near extreme black hole. We will investigate this
problem in a separate work \cite{bcde08b}.

We also presented an analytic solution for the massless scalar
field in the metric of the Kerr-Newman naked singularity, i.~e.
when $M^2<Q^2+a^2$. This solution describes the scalar field
distribution around the naked singularity, in general with
non-zero influx. The found solution contains some pathologies
which are the consequences of the causality violation in the
vicinity of the Kerr-Newman space-time with a naked singularity.
For the Kerr or the Reissner-Nordstr\"om naked singularity the
presented solution is well behaved in the whole spacetime, except
for the point of the physical singularity.

\begin{acknowledgments}
We would like to thank V.~Beskin, Ya.~Istomin, V.~Lukash, and
K.~Zybin for useful discussions. The work of EB was supported by
the EU FP6 Marie Curie Research and Training Network
``UniverseNet'' (MRTN-CT-2006-035863). The work of other coauthors
was supported in part by the Russian Foundation for Basic Research
grant 06-02-16342 and the Russian Ministry of Science grant LSS
959.2008.2.
\end{acknowledgments}

\appendix
\section{Alternative formalism for accretion}
\label{sec rest}
Here we present an alternative approach to solve the problem of
accretion in a specific case when a black hole is at rest,
$u^0_\infty=1$, $u_\infty=0$. The way we solve the problem here is
closer to the original ``hydrodynamic'' approach, described in
\cite{bondi,zeldovich,Michel}, rather that the ``scalar field
approach'' by Pertich {\sl et al} \cite{PetShapTeu}.

The equations for stationary distribution of ultra-hard fluid with
the equation of state $p=\rho$ in the Kerr-Newman metric is
integrated directly, similar to the analogous problem in the
Schwarzschild \cite{Michel,bde04,bde05} and Reissner-Norestr\"om
metrics \cite{bcde08a}. From (\ref{4velocity}) it follows that the
specific azimuthal and longitudinal angular momentum are both
zero:
\begin{equation}
 L_\phi=u_\phi=0, \quad L_\theta=u_\theta=u^\theta=0.
\end{equation}
This means, in particular, that $u^\theta=0$, and so
$\theta=const$ along the lines of flow. Using this property of the
stationary ultra-hard fluid, we find the first integrals of the
energy momentum conservation
\begin{equation}
 T^\alpha_{\beta\,;\alpha}=\frac{1}{\sqrt{-g}}
 \frac{\partial}{\partial x^\alpha}(T^\alpha_\beta\sqrt{-g})
 =\frac{1}{2}g_{\alpha\gamma,\beta}T^{\alpha\gamma}=0.
\end{equation}
Integration of this equation for $\beta=0$ gives the first
integral of motion (the relativistic Bernoulli energy conservation
equation):
\begin{equation}
 (p+\rho)u_0u\sqrt{-g}=C_1(\theta)M^2,
 \label{integral1}
\end{equation}
where $u=u^r$ is a radial 4-velocity component and $C_1(\theta)$
is a function of $\theta$. To find the second integral of motion
we write a projection equation
\begin{equation}
 u_\mu T^{\mu\nu}_{;\nu}=\rho_{,\nu}u^\nu+(p+\rho)\frac{1}{\sqrt{-g}}
 \frac{\partial}{\partial
 x^\alpha}\left(\sqrt{-g}u^\alpha\right)=0,
\end{equation}
which can be expressed as
\begin{equation}
 \frac{\rho_{,r}}{p+\rho}u+
 \frac{1}{\sqrt{-g}}\frac{\partial}{\partial r}(\sqrt{-g}u)=0.
 \label{projection}
\end{equation}
Integration of (\ref{projection}) gives the second integral of
motion (the energy flux conservation):
\begin{equation}
 \sqrt{-g}u\exp\left[\,\,\int\limits_{\rho_{\infty}}^{\rho}
 \frac{d\rho'}{\rho'+p(\rho')}\right]=-A(\theta)M^2,
 \label{integral2}
\end{equation}
where $A(\theta)$ is a function of $\theta$. Using normalization
condition $u^\alpha u_\alpha=1$, from integrals of motion
(\ref{integral1}) and (\ref{integral2}) we find
\begin{eqnarray}
 &&(p\!+\!\rho)\exp\!\left[-\!\!\int\limits_{\rho_{\infty}}^{\rho}
 \frac{d\rho'}{\rho'+p(\rho')}\right]\!
 \!\!\sqrt{\frac{\Sigma(\Delta+\Sigma u^2)}{\mathcal A}}
 \!=\!-\frac{C_1(\theta)}{A(\theta)} \nonumber \\
 &&=p_\infty+\rho_\infty,
 \label{integral2b}
\end{eqnarray}
where function $\mathcal A$ is given by (\ref{A}). From
Eqs.~(\ref{integral2}) and (\ref{integral2b}) for $a\to0$ and
$Q\to0$, we can fix the functions $C_1(\theta)$ and $A(\theta)$,
$$C_1(\theta)=-A(\theta)(p_\infty+\rho_\infty), \quad
A(\theta)=A_0\sin\theta,$$ where $A_0=const$ is a dimensionless
constant. Following Michel \cite{Michel}, we find relations at the
critical sound point $r=r_*$:
\begin{equation}
 \Delta(r_*)\!=\!0, \:\:
 u_\ast^2\!=\!\frac{(r_*^2+a^2)^2(r_*-M)}{\Sigma[2r_*(r_*^2\!+\!a^2)
 \!-\!a^2\sin^2\theta(r_*\!-\!M)]},
 \label{critical}
\end{equation}
where $u_*=u(r_*)$. From the first equation in (\ref{critical}) it
follows that for $M^2\geq a^2+Q^2$, there are two critical points
$r_1=r_-$ and $r_2=r_+$. The critical point at smaller radius,
$r_1=r_-$, is inside the event horizon. The two points coincide
only for the extreme black hole, $M^2=a^2+Q^2$. This means that
for the extreme black hole the boundary conditions must be
different from ones in the non-extreme case. Using the general
relation $b=2+i-s$, derived in \cite{Beskin97,Beskin04,Bogovalov},
between the number of boundary conditions $b$, the number of
invariants $i$ and the number of critical surfaces $s$, we can see
that in the extreme case we must put only 3 boundary conditions,
very similar to the case of a stationary atmosphere. Thus, a
stationary atmosphere seems to be a more adequate description in
the case of the extreme black hole, rather than a stationary
accretion.

Using the relations (\ref{critical}) for critical point $r=r_+$,
we calculate the constant $A_0$, which determines the value of the
accretion flow:
\begin{equation}
 A_0=\frac{r_+^2+a^2}{M^2}.
\end{equation}
Now, from (\ref{integral2}) and (\ref{integral2b}) one can find
the space distribution of the energy density,
$\rho=\rho(r,\theta)$, and the radial component of the 4-velocity,
$u=(A_0M^2/\Sigma)(\rho/\rho_\infty)^{-1/2}$,
\begin{eqnarray}
\label{solution again}
 \frac{\rho}{\rho_\infty} &\!\!=&\!\!
 \frac{(r\!+\!r_+)(r^2\!+\!r_+^2\!+\!2a^2)\!-\!a^2(r\!
 -\!r_-)\sin^2\theta}{\Sigma(r-r_-)},\\
 u^2&\!\!=&\!\!\frac{(r_+^2+a^2)^2(r-r_-)}{\Sigma[(r\!+\!r_+)(r^2\!
 +\!r_+^2\!+\!2a^2)\!-\!a^2(r\!-\!r_-)\sin^2\theta]}. \nonumber
\end{eqnarray}
The solution (\ref{solution again}) coincides with the one found
in Sec.~\ref{sec moving} in the case when $u_\infty=0$.


\begin{thebibliography}{99}

\bibitem{PetShapTeu} L.~Petrich, S.~Shapiro and S.~Teukolsky,
Phys. Rev. Lett. {\bf 60}, 1781 (1988).

\bibitem{shapiro89} S. L.~Shapiro, Phys. Rev. {\bf D39}, 2839 (1989).

\bibitem{psst89} L.~Petrich, S. L.~Shapiro, R. F.~Stark and
S.~Teukolsky, Astrophys. J. {\bf 336}, 313 (1989).

\bibitem{AbrahamsShapiro90} A. M.~Abrahams and S. L.~Shapiro,
Phys. Rev. {\bf D41}, 327 (1990).

\bibitem{KarasMucha} V.~Karas and R.~Mucha, Am.~J.~Phys. {\bf 61},
825 (1993).

\bibitem{LLfluid} L. D.~Landau and E. M.~Lifshitz, {\it Fluid
Mechanics}, (Addison and Wesley Rading MA, 1959).

\bibitem{moncrief} V.~Moncrief, Astrophys. J. {\bf 235}, 1038
(1980).

\bibitem{mtw} C. W. Misner, K.S. Thorne and J. A. Wheeler, {\it
Gravitation}, (W. H. Freeman and Co, San Francisco, 1973), Chapter
33.

\bibitem{Gal'tsov} D. M. Gal'tsov, {\it Particles and fields
in the vicinity of black holes Gravitation}, (Moscow, Izdatel'stvo
Moskovskogo Universiteta, 1986) in Russian.

\bibitem{bondi} H. Bondi, Mon. Not. Roy. Astron Soc {\bf 112}, 195
(1952).

\bibitem{zeldovich} Ya. B. Zeldovich and I. D. Novikov, {\it
Relativistic astrophysics. Vol.1: Stars and relativity}, (Chicago:
University of Chicago Press, 1971), Chapter 12.

\bibitem{Michel} F. C.~Michel, {Astrophys. Space Sci.} {\bf 15},
153 (1972).

\bibitem{Beskin97} V.~Beskin, Physics-Uspekhi {\bf 40}, 659 (1997).

\bibitem{Beskin04} V.~S.~Beskin, {\it Astrophysics and Cosmology After
Gamow}, edited by G. S. Bisnovaty-Kogan, S. Silich, E. Terlevich, R.
Terlevich and A. Zhuk, (Cambridge Scientific Publishers, Cambridge, UK,
2007), p.201

\bibitem{Bogovalov} S. V.~Bogovalov, Astron. Astrophys. {\bf 323}
634 (1997).

\bibitem{abram} M.~Abramowitz and I. A.~Stegun, {\it Handbook of
Mathematical Functions}, (New York: Dover, 1972), Eq. (8.1.2).

\bibitem{PolZai} A. D. Polaynin and V. F. Zaisev, {\it Handbook of
exact solutions for ordinary differential equations}, (CRC Press,
Boca Raton, 1995).

\bibitem{carter68} B. Carter, Phys.Rev. {\bf 174}, 1559 (1968).

\bibitem{bardeen1} J. M.~Bardeen, Astrophys. J. {\bf 162}, 71 (1970).

\bibitem{bardeen2} J. M.~Bardeen, W. H.~Press and S. A.~Teukolsky,
Astrophys. J. {\bf 178} 347 (1972).

\bibitem{bde04} E.~Babichev, V.~Dokuchaev and Yu.~Eroshenko,
Phys. Rev. Lett. {\bf 93} 021102 (2004).

\bibitem{bde05} E.~Babichev, V.~Dokuchaev and Yu.~Eroshenko,
Zh. Eksp. Teor. Fiz. {\bf 127}, 597 (2005) [JETP {\bf 100}, 528
(2005) (translated from Russian)].

\bibitem{bcde08a} E.~Babichev, S.~Chernov, V.~Dokuchaev and
Yu.~Eroshenko, arXiv:0806.0916 [gr-qc].

\bibitem{hod02} S.~Hod, Phys. Rev. {\bf D66}, 024016 (2002).

\bibitem{hod08} S.~Hod, Phys. Rev. Lett. {\bf 100}, 121101 (2008).

\bibitem{bcde08b} E.~Babichev, S.~Chernov, V.~Dokuchaev and
Yu.~Eroshenko, {\it in preparation}.

\end{thebibliography}
\end{document}